\documentclass{webofc}
\usepackage[varg]{txfonts}   
\usepackage{amsmath,bm}

\long\def\comment#1{ }
\newcommand{\eqn}[1]{Eq.~\eqref{#1}}
\newcommand{\beq}{\begin{equation}}
\newcommand{\eeq}{\end{equation}}
\newcommand{\nn}{\nonumber\\}
\newcommand{\dif}{{\rm d}}

\newcommand{\rme}{{\rm e}}
\newcommand{\rmi}{{\rm i}}
\newcommand{\rmP}{{\rm P}}

\newcommand{\rmtr}{{\rm tr}}

\newcommand{\mcal}{\mathcal}

\newcommand{\bk}{\bm{k}}

\newcommand{\bx}{\bm{x}}
\newcommand{\by}{\bm{y}}

\newcommand{\bz}{\bm{z}}

\newcommand{\br}{\bm{r}}

\newcommand{\abar}{\bar{\alpha}_s}

\begin{document}

\title{Forward particle production in proton-nucleus collisions at next-to-leading order}

\author{\firstname{D.N.} \lastname{Triantafyllopoulos}\inst{1}\fnsep\thanks{\email{trianta@ectstar.eu}}}

\institute{European Centre for Theoretical Studies in Nuclear Physics and Related Areas (ECT*) \\
and Fondazione Bruno Kessler, Strada delle Tabarelle 286, I-38123 Villazzano (TN), Italy}

\abstract{We consider the next-to-leading order (NLO) calculation of single inclusive particle production at forward rapidities in proton-nucleus collisions and in the framework of the Color Glass Condensate (CGC). We focus on the quark channel and the corrections associated with the impact factor. In the first step of the evolution the kinematics of the emitted gluon is kept exactly (and not in the eikonal approximation), but such a treatment which includes NLO corrections is not explicitly separated from the high energy evolution. Thus, in this newly established ``factorization scheme'', there is no ``rapidity subtraction''. The latter suffers from fine tuning issues and eventually leads to an unphysical (negative) cross section. On the contrary, our reorganization of the perturbation theory leads by definition to a well-defined cross section and the numerical evaluation of the NLO correction is shown to have the correct size.} 

\maketitle

\section{Introduction}
\label{intro}

When doing perturbation theory in Quantum Mechanics, e.g.~when calculating the energy change in the states of a system due to a small perturbation, normally it suffices to sum a few terms. That is, $E_n = \sum_{k=0}^{k_0} g^{2k}E_n^{(k)}$, with $g^2$ the size of the perturbation, $E_n^{(k)}$ fixed (as they depend only on a few  parameters of the unperturbed Hamiltonian) and $k_0$ an integer chosen according to the desired accuracy. In an interacting Quantum Field Theory, like QCD, very often one is interested in a cross section, which perturbatively reads $\sigma = \sigma_0 + g^2 \sigma_1 +g^4 \sigma_2 \cdots$. Now things can be different since the coefficients $\sigma_k$ can also depend on the momenta of the outgoing particles: although the QCD coupling $\alpha_s = g^2/4\pi$ is small at short distances, for some momenta one may have $\alpha_s \sigma_{k+1} \sim \sigma_{k}$. Then the fixed order series does not provide accurate information and one must resum to all orders in $\alpha_s$ in such kinematical domains. 

\begin{figure}[t]
\begin{center}
\begin{minipage}[b]{0.40\textwidth}
\begin{center}
\includegraphics[width=0.6\textwidth,angle=0]{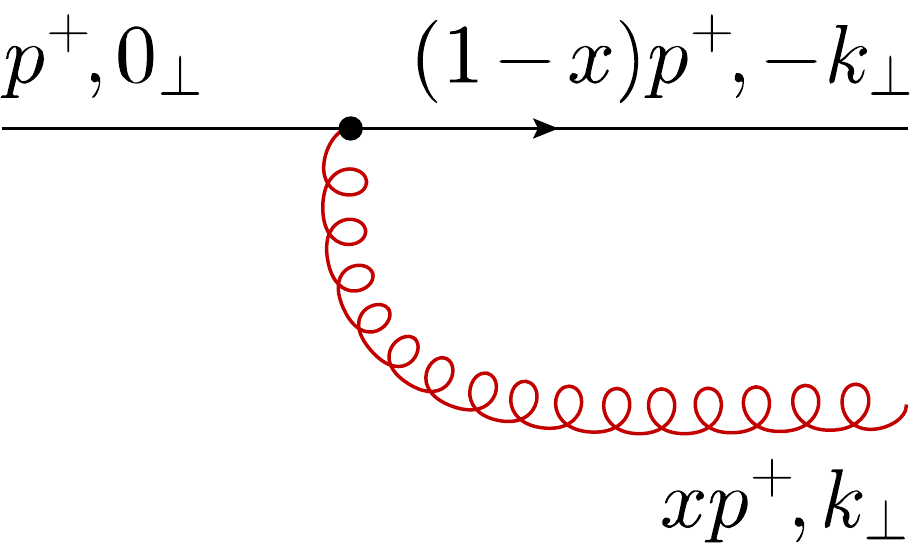}\\(a)\vspace{0cm}
\end{center}
\end{minipage}
\hspace{0.05\textwidth}
\begin{minipage}[b]{0.40\textwidth}
\begin{center}
\includegraphics[width=0.65\textwidth,angle=0]{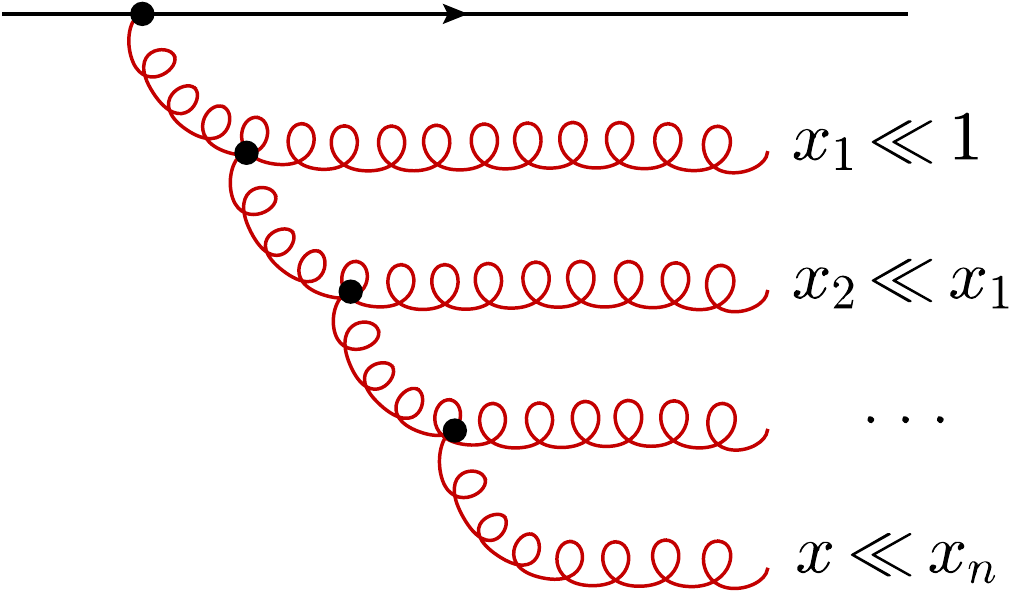}\\(b)\vspace{0cm}
\end{center}
\end{minipage}
\end{center}
\vspace{-0.25cm}
\caption{(a) Single gluon emission from a parton and (b) strongly ordered gluon emissions.}
\label{fig-dp}
\end{figure}

Even one of the most basic emissions in QCD, that of a gluon from a parent parton (quark or gluon), exhibits such features. Starting from the amplitude in Fig.~\ref{fig-dp}.a, we find that the differential probability under consideration in the double logarithmic limit\footnote{The logarithmic limit in transverse momentum is taken just for simplicity and in fact should be relaxed.} reads 
\begin{equation}
	\dif P = \frac{\alpha_s C_{\rm R}}{\pi}\,
	\frac{\dif k_{\perp}^2}{k_{\perp}^2}\,\frac{\dif x}{x},
\end{equation}     
with $C_{\rm R}$ the Casimir for the representation of the parent parton. Say we want to measure the number of gluons with a very small longitudinal momentum fraction $x$. Then, at order $n+1$ in perturbation theory all intermediate gluons with strongly ordered longitudinal momenta as in Fig.~\ref{fig-dp}.b are integrated over, leading to a large ``coefficient'' proportional to $\ln^{n}(1/x)$. When $x$ is sufficiently small, this coefficient becomes comparable to (or larger than) $\alpha_s^n$, and twe must sum over all $n$, i.e.~over all the diagrams with an arbitrary number of strongly ordered gluons. We find that the result exponentiates and, for example, the integrated gluon distribution (the number of gluons per unit rapidity and with transverse momentum less than $Q^2$) grows like
\begin{equation}
	xG(x,Q^2) \sim \frac{1}{x^{\lambda}} 
	\quad \textrm{with} \quad \lambda = 0.2 \div 0.3.
\end{equation}

Such a growth cannot continue to arbitrary small-$x$, since gluons start to overlap and the interaction among them becomes strong, even though the coupling is still weak at the transverse scale of interest. ``Gluon saturation'' will occur at a scale for which the number of gluons per unit phase space becomes of order $1/\alpha_s$, that is when $xG(x,Q_s^2)/Q_s^2 \pi R^2 \sim 1/\alpha_s$, where $R$ is the size of the hadron under study. Then the saturation momentum for a large nucleus of atomic number $A$ reads  
\begin{equation}
	Q_s^2(x) \sim Q_0^2 A^{1/3} x^{-\lambda_s},
\end{equation}
with $Q_0^2$ being non-perturbative, while $\lambda_s$ can be determined perturbatively. For small-$x$ and/or large $A$, the saturation scale is much larger than $\Lambda_{\rm QCD}$ and indeed weak coupling techniques are applicable. Still, apart from the resummation of large logarithms, one eventually needs to deal with non-linear dynamics due to saturation. A QCD effective theory for describing such phenomena is the CGC \cite{Gelis:2010nm}.

The soft part of the hadronic wavefunction can be probed, for example, in proton-nucleus collisions in the forward region of the proton. A quark collinear with the proton and with a moderate to large momentum fraction $x_p = p^+/Q^+$, multiply scatters with the soft gluons of the nucleus that carry a fraction $X_g=p^-/P^-$, and picks up a transverse momentum $k_{\perp}$ so that it moves at an angle $\theta$ with the collision axis. Here, $Q^+$ is the longitudinal light-cone momentum of the proton which moves in the positive direction, while $P^-$ is that of the nucleus moving towards the negative one. A simple kinematics exercise shows that the quark rapidity in the COM reads $\eta=-\ln\tan(\theta/2)$, while $x_p= k_{\perp} \rme^{\eta}/\sqrt{s}$ and $X_g= k_{\perp} \rme^{-\eta}/\sqrt{s}$, with $s=2 Q^+ P^-$ the COM energy squared. Then it is obvious that the forward region $\theta \ll 1$ maps to small-$X_g$. We also note that in the infinite momentum frame of the nucleus, which shall be used later on, one has $X_g = k_{\perp}^2/x_p s$.

Resumming both $[\abar \ln(1/X_g)]^n$ and $\abar[\abar \ln(1/X_g)]^n$ terms, with $\abar = \alpha_s N_c/\pi$ and $N_c$ the number of colors, in the presence of saturation, one has been able to give the cross section for single inclusive particle production at NLO in the CGC framework \cite{Chirilli:2011km,Chirilli:2012jd}. The problem is that such a cross section when evaluated numerically turns out to be negative (although it is positive at LO) for values $k_{\perp} \gtrsim Q_s$ \cite{Stasto:2013cha}. Here we shall review the process under consideration, and provide for the solution in the aforementioned problem \cite{Iancu:2016vyg}.

\section{Particle production at LO and the BK equation}
\label{sec-lo}

\begin{figure}[t]
\begin{center}
\begin{minipage}[b]{0.5\textwidth}
\begin{center}
\includegraphics[width=1.1\textwidth,angle=0]{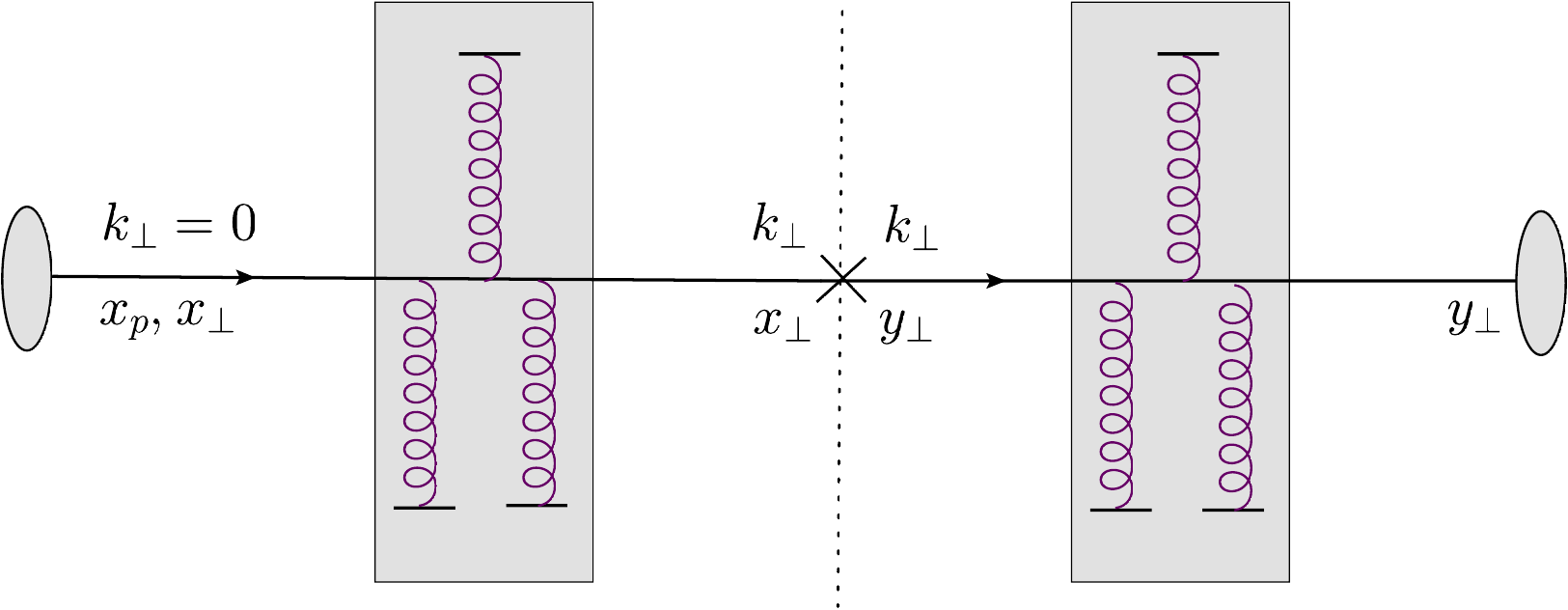}\\(a)\vspace{0cm}
\end{center}
\end{minipage}
\hspace{0.05\textwidth}
\begin{minipage}[b]{0.4\textwidth}
\begin{center}
\includegraphics[width=0.9\textwidth,angle=0]{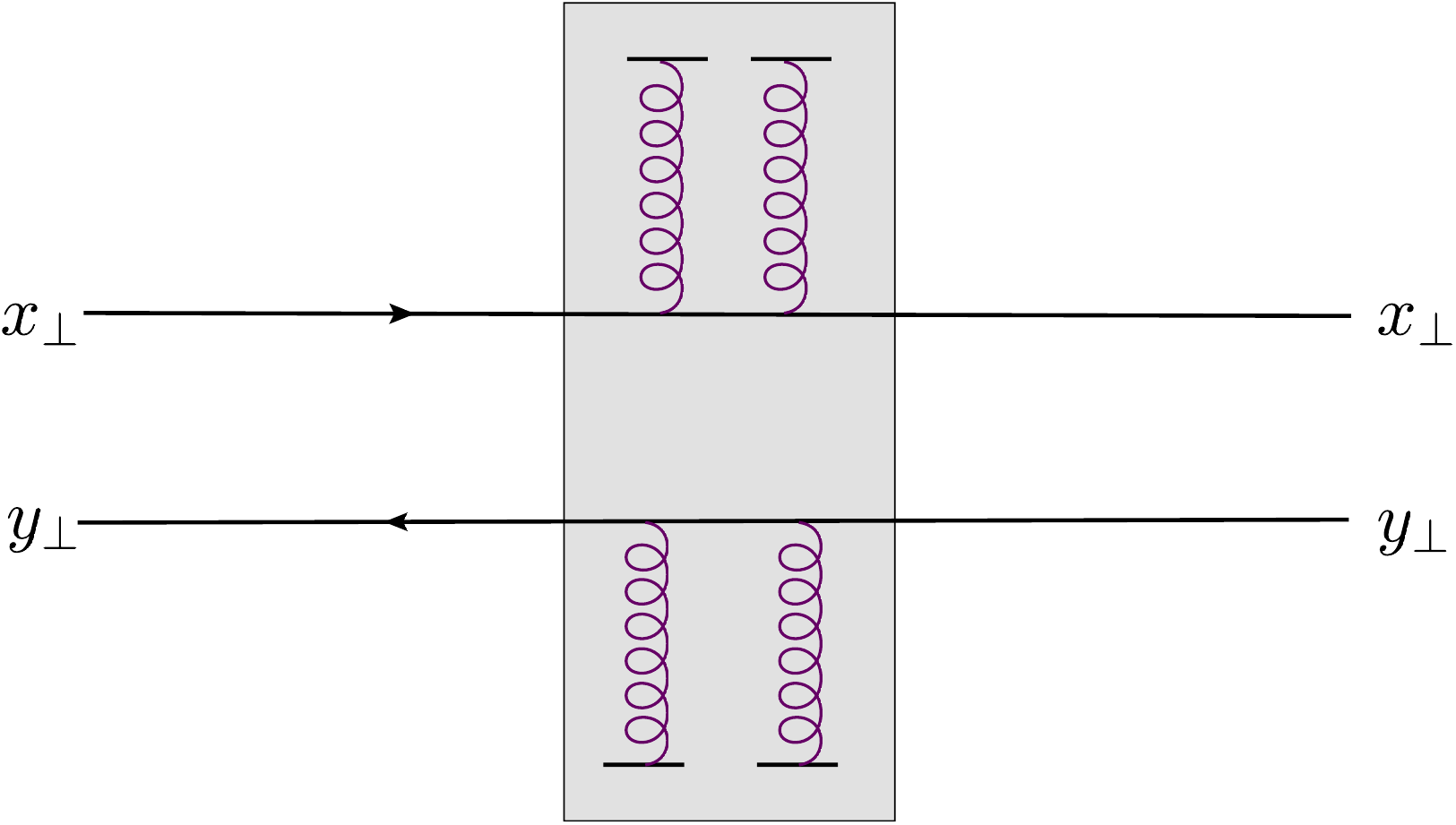}\\(b)\vspace{0cm}
\end{center}
\end{minipage}
\end{center}
\vspace{-0.25cm}
\caption{(a) Quark scattering off the nucleus in the CA and DDA and (b) dipole-nucleus scattering.}
\label{fig-mult}
\end{figure}

In our analysis we shall omit the fragmentation of the outgoing forward quark into hadrons, since it does not play any role in the problem under discussion and as it is anyway straightforward to insert it at any time in the calculation. At LO, the collinear with the proton quark scatters multiply off the target field $A^{\mu}$, which can be strong for transverse momenta around (or smaller than) $Q_s$. The quark moves having its transverse coordinate fixed at $\bx$, while its color undergoes a precession and such an eikonal interaction is given by the Wilson line 
\begin{equation}
	V(\bx) = \rmP \exp \left[\rmi g \int \dif x^+ A_a^-(x^+,\bx) t^a \right],
\end{equation}
where P stands for path ordering along the light-cone time $x^+$ of the quark and $t^a$ is in the fundamental representation. The direct amplitude (DA) $\mathcal{M}_{ij}(\bk)$ is just the two-dimensional Fourier transform of $V_{ij}$ and then the single inclusive production for a particle (here the quark) with transverse momentum $\bk$ and rapidity $\eta$ is obtained by multiplying with the complex conjugate amplitude (CCA) (see Fig.~\ref{fig-mult}.a) and by averaging/summing over colors. It reads
\begin{equation}
\label{dnlo}
	\frac{\dif N^{\rm LO}}{\dif \eta \dif^2 \bk} =
	\frac{x_{p} q(x_p)}{(2\pi)^2} \int \dif^2\br\,	
	\rme^{-\rmi \bk \cdot \br} S(\br,X_g), 
\end{equation}
where $x_p q(x_p)$ is the quark pdf in the proton, $\br=\bx-\by$ and $S(\br;X_g)$ is the elastic $S$-matrix for the $q\bar{q}$ dipole ($\bx,\by$) to scatter off the nucleus\footnote{In fact such an expression has been proved to be valid even at NLO \cite{Mueller:2012bn}.} (see Fig.~\ref{fig-mult}.b). This $S$-matrix is bilinear in the Wilson lines (as we have taken the product of the DA with the CCA), and is given by
\begin{equation}
\label{svv}
	S(\bx,\by,X_g) = \frac{1}{N_c}
	\Big\langle 
	\rmtr\big[ V(\bx) V^{\dagger}(\by)\big]
	\Big\rangle_{X_g}.
\end{equation}
It is evaluated at the target scale $X_g$, which is determined from the kinematics, and resums the aforementioned powers of $\abar \ln (1/X_g)$. In the problem at hand, we have already assumed in \eqn{dnlo} that the nucleus is homogeneous so that $S$ does not depend on the impact parameter $\bm{b} = (\bx+\by)/2$. When expanding the gauge field $A^{\mu}$ to second order, i.e.~when neglecting the multiple scattering, the cross section becomes proportional to the nuclear gluon pdf as it should. We do not elaborate on the details of the averaging in \eqn{svv}, but it suffices to say that it amounts to specifying the initial condition of an equation that we now discuss.

\begin{figure}[t]
\begin{center}
\includegraphics[width=0.6\textwidth,angle=0]{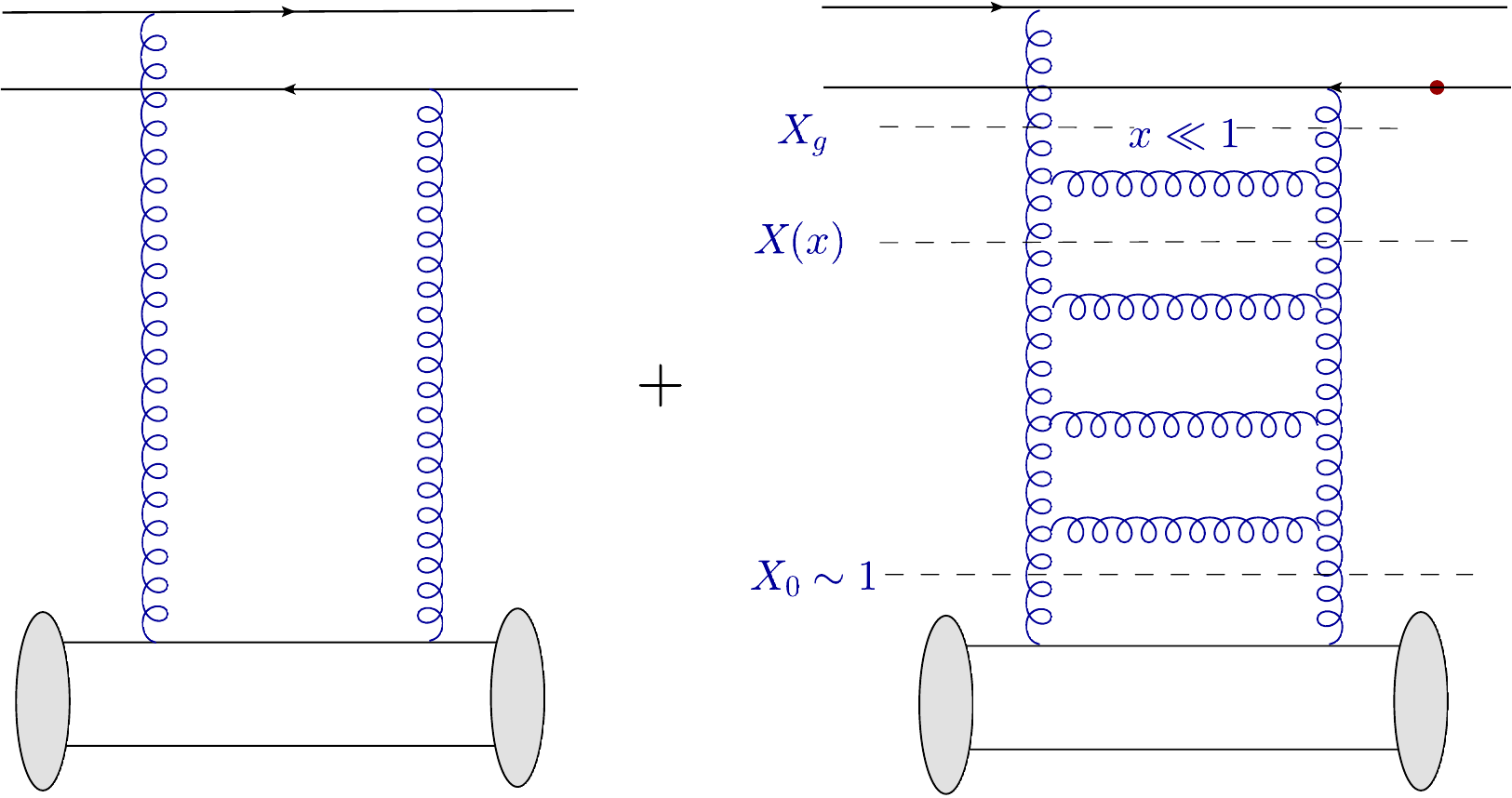}\vspace{-0.25cm}
\end{center}
\caption{Particle production at LO.}
\label{fig-lo}
\end{figure}

The easiest way to implement the resummation of the large logarithms is to write the BK equation \cite{Balitsky:1995ub,Kovchegov:1999yj} which describes the evolution of the elastic dipole $S$-matrix. For our purposes, it is more convenient to present such an equation in its integral form which reads
\begin{equation}
\label{bk}
	S(\br,X_g) = S(\br,X_0) + \frac{\abar}{2\pi} 
	\int_{X_g}^{1} \frac{\dif x}{x} \int \dif^2 \bz \,
	\frac{\br^2}{\bz^2 (\br-\bz)^2}
		[S(\bz) S(\br-\bz) -S(\br)]_{X(x)},
\end{equation}
where $X(x)=X_g/x$ is the longitudinal momentum fraction in the target in the arbitrary intermediate evolution step. The kernel in \eqn{bk} is called the dipole kernel \cite{Mueller:1993rr} as it stands for the transverse dependence of the probability for the dipole $\br$ to split into $\bz$ and $\br-\bz$. $S(\br,X_0)$ is the tree level contribution, the one in the absence of any QCD evolution, and typically it is given by the McLerran-Venugopalan (MV) model \cite{McLerran:1993ka} or some variant of it. Such an equation is better understood in a frame where almost all of the evolution is associated with the nucleus, apart from the emission of the primary gluon (the one closest to the dipole $\br$), cf.~Fig.~\ref{fig-lo}. It is clear that the solution to Eq.~\eqref{bk} unitarizes, i.e.~$S\to 0$ when $X_g \to 0$ for a fixed dipole size $r_{\perp} = |\br|$ (as we assume that $S$ depends only on the magnitude of $\br$). In the current setup, it is natural to define the saturation momentum as the line in the $r_{\perp}$-$X_g$ plane along which the $S$-matrix is fixed to a value of order (but of course smaller than) 1, e.g.
\begin{equation}
	S(r_{\perp}=2/Q_s(X_g),X_g) = 1/2.
\end{equation}
Indeed one finds that $Q_s^2(X_g)$ grows as an inverse power of $X_g$, although the growth is too strong. Using NLO BK dynamics and resumming collinear logarithms which can get large, the growth is tamed \cite{Triantafyllopoulos:2002nz,Beuf:2014uia,Iancu:2015vea} and comes to a good agreement with the phenomenology. Moreover, eventually one is able to reasonably describe the relevant data in d-Au collisions using Eq.~\eqref{bk} with running coupling and inserting $K$-factors to account for the remaining higher order corrections \cite{Dumitru:2005gt,Albacete:2012xq}.

\section{Particle production at NLO}
\label{sec-nlo}

\begin{figure}[t]
\begin{center}
\begin{minipage}[b]{0.42\textwidth}
\begin{center}
\includegraphics[width=0.98\textwidth,angle=0]{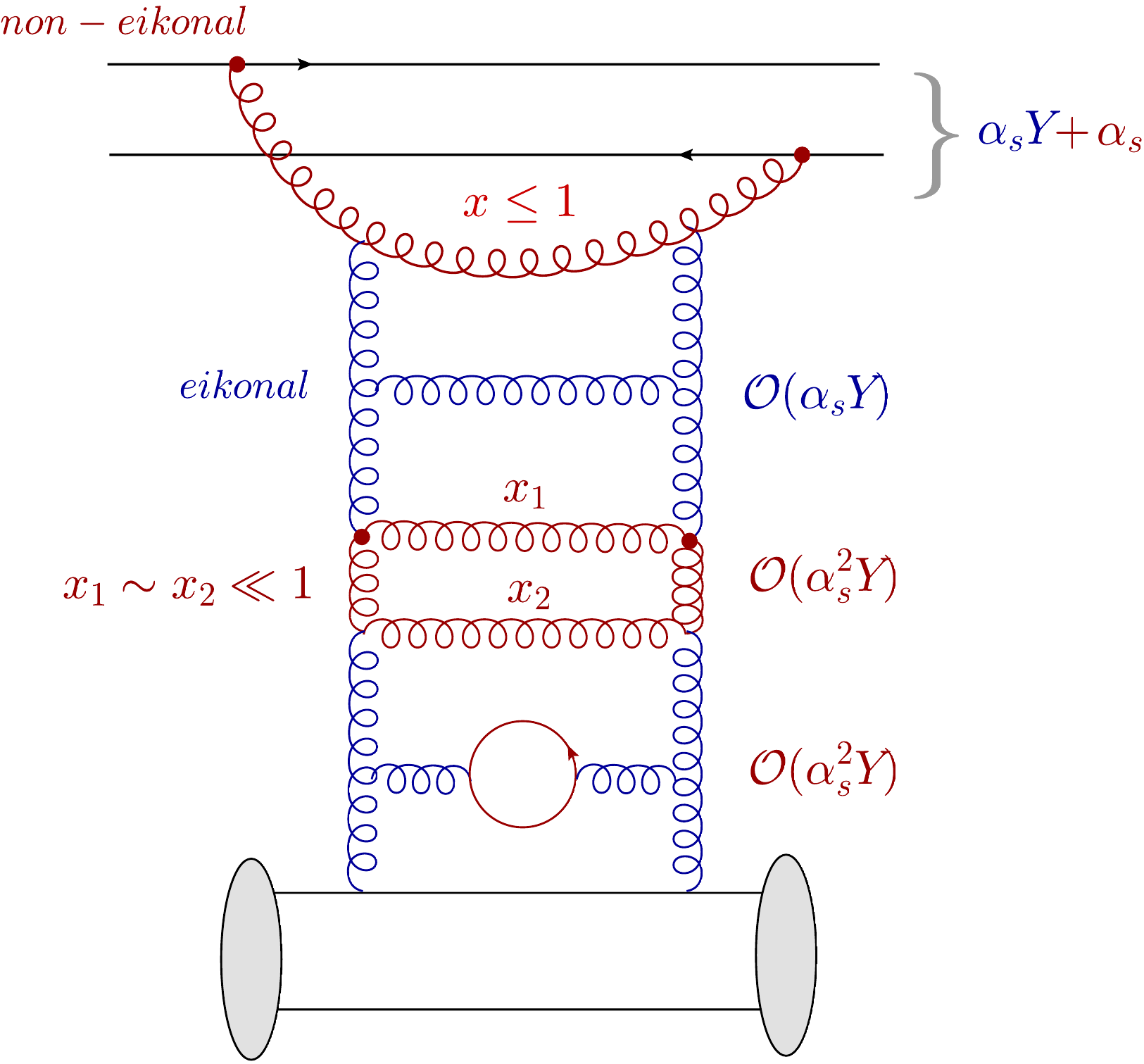}\\(a)\vspace{0cm}
\end{center}
\end{minipage}
\hspace{0.03\textwidth}
\begin{minipage}[b]{0.5\textwidth}
\begin{center}
\includegraphics[width=1.12\textwidth,angle=0]{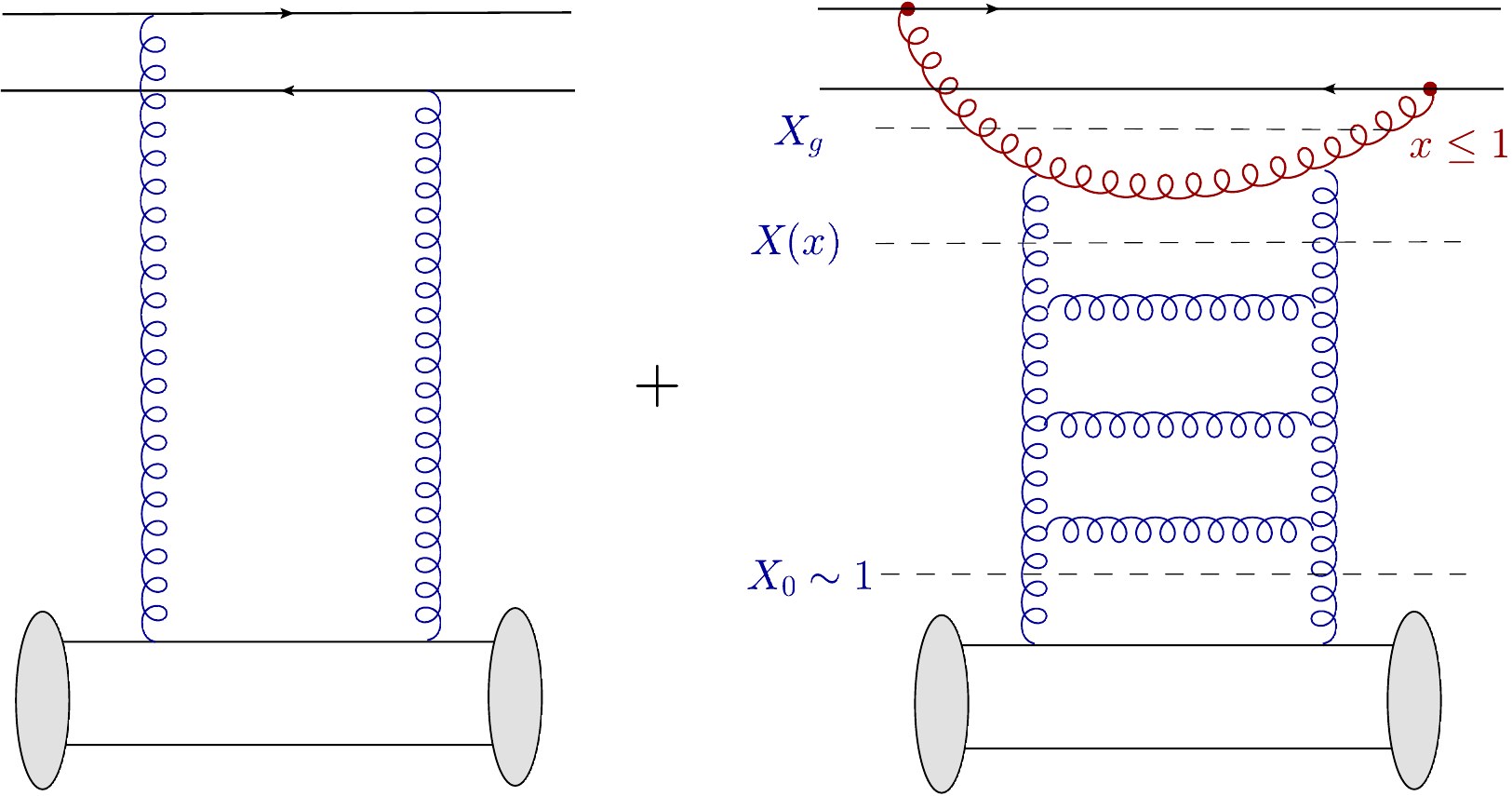}\\(b)\vspace{0cm}
\end{center}
\end{minipage}
\end{center}
\vspace{-0.25cm}
\caption{(a) NLO corrections and (b) NLO factorization.}
\label{fig-nlo}
\end{figure}

As depicted in Fig.~\ref{fig-nlo}.a, NLO corrections to the single inclusive particle production, can be separated in two parts. First, one may have two adjacent gluons in the ``ladder'' which have fractions such that $x_n \sim x_{n+1} \ll 1$, or one may have a loop. These are corrections associated with the evolution of the color dipole and can be taken into account by using the NLO BK equation. We shall not deal with them, since in principle they can be implemented after collinear resummations have been done without creating any additional issues. Second, one has a NLO correction when the primary gluon is not emitted eikonally, i.e.~when its fraction $x$ is of order (but smaller than) 1. Thus, by treating exactly the kinematics one can include this impact factor correction (still, one must notice that the scattering with the nucleus remains eikonal and is given by Wilson lines) into the calculation of the cross section. Then, by extending the LO result in Eqs.~\eqref{dnlo} and \eqref{bk}, it is not hard to see that at NLO we have the sum of the two terms in Fig.~\ref{fig-nlo}.b. In fact there are two NLO contributions, one proportional to $N_c$ and one to $C_{\rm F}$, and we focus on the former which reads
\begin{align}
	\label{dnnlonc}
	\frac{\dif N}{\dif\eta\dif^2\bk} = &
	\frac{x_{p} q(x_p)}{(2\pi)^2} \mcal{S}(\bk,X_0) + 
	\frac{1}{4\pi}\int_0^{1-X_g/X_0} \dif \xi\,
	\frac{1 +\xi^2}{1-\xi}
	\nn
	&\times
	\left[ \Theta(\xi - x_p)\frac{x_p}{\xi}
	\mathcal{J}\big(\bk,\xi,X(\xi)\big)
	- x_p q(x_p) 
	\mathcal{J}_{v}\big(\bk,\xi,X(\xi)\big)
	\right].  
\end{align}
The real $\mathcal{J}$ and virtual $\mathcal{J}_v$ pieces (implicitly proportional to $\abar$) are known \cite{Chirilli:2011km,Chirilli:2012jd,Ducloue:2016shw}, $\mcal{S}$ stands for the Fourier transform of $S$, and $1-\xi = x$ is the fraction of the primary gluon. In a compact notation, where we omit the quark pdf, suppress the transverse coordinates and trivial prefactors, and portray collectively the scattering of the $q\bar{q}g$ system, we can write 
\begin{equation}
\label{dnnlo}
	\frac{\dif N}{\dif\eta\dif^2\bk} = 
	\mcal{S}_0(\bk) + \abar \int_{X_g}^{1} \frac{\dif x }{x}\,
	\mcal{K}(x)\,\mcal{S}_{q\bar{q}g}\big(\bk,X(x)\big).
\end{equation}
\eqn{dnnlo} is our main result and being the sum of the tree level contribution and a positive term should not pose any problems if no further approximations are done \cite{Iancu:2016vyg}. Indeed, this will be manifest in the numerical results to be shown in the next section.

\section{$k_{\perp}$-factorization: fine-tuning and negativity issues}
\label{sec-kperp}

\eqn{dnnlo} is not in the traditional form of $k_{\perp}$-factorization. This is so, because the kernel $\mcal{K}$ depends not only on the transverse coordinates, but also on $x$, and moreover, the $S$-matrices of the $q\bar{q}g$ system are evaluated at the ``floating'' scale $X(x)$. Now the task is to see how we can arrive at a $k_{\perp}$-factorized expression. First notice that the LO result is recovered, as it should, by letting in \eqn{dnnlo} $\mcal{K}(x) \to \mcal{K}(0)$: indeed, the two terms in the r.h.s.~combine to give $S(\bk,X_g)$. Then, we are free to add and subtract this LO piece to the r.h.s.~of \eqn{dnnlo} and by doing so we get the ``subtracted'' expression
\begin{equation}
\label{dnnlo2}
	\frac{\dif N}{\dif\eta\dif^2\bk}	 = \mcal{S}(\bk,X_g) 
	+ \abar \int_{X_g}^{1} \frac{\dif x }{x}\,
	\big[\mcal{K}(x) - \mcal{K}(0)\big]\,\mcal{S}_{q\bar{q}g}\big(\bk,X(x)\big).
\end{equation}
This is a mathematically correct step, however it is a dangerous one, since there is reshuffling of a very large contribution between the two terms. In order to obtain the correct result from \eqn{dnnlo2}, one must be extremely precise on the numerical evaluation of the two terms \cite{Iancu:2016vyg}.

Two more steps, which now involve approximations, are required to bring \eqn{dnnlo2} in a $k_{\perp}$-factorized form. Since $\mcal{K}(x) - \mcal{K}(0)$ clearly vanishes as $x\to 0$, the integrand on the r.h.s.~is not anymore dominated by small values of $x$, and therefore at NLO accuracy we can let $\mcal{S}_{q\bar{q}g}\big(\bk,X(x)\big) \to \mcal{S}_{q\bar{q}g}\big(\bk,X_g\big)$ and furthermore let $X_g \to 0$ in the lower limit of the integration. Then we arrive at \cite{Iancu:2016vyg} 
\begin{equation}
\label{dnnlo3}
	\frac{\dif N^{k_{\perp-{\rm fact}}}}{\dif\eta\dif^2\bk}	 = \mcal{S}(\bk,X_g) 
	+ \abar \int_{0}^{1} \frac{\dif x }{x}\,
	\big[\mcal{K}(x) - \mcal{K}(0)\big]\,\mcal{S}_{q\bar{q}g}\big(\bk,X_g\big),
\end{equation}
which is what ``standard'' $k_{\perp}$-factorization would give: the $S$-matrices are evaluated locally at $X_g$, while the kernel depends only in the transverse coordinates (since, in principle, we can now integrate over $x$). 

\begin{figure}[t]
\begin{center}
\begin{minipage}[b]{0.45\textwidth}
\begin{center}
\includegraphics[width=0.95\textwidth,angle=0]{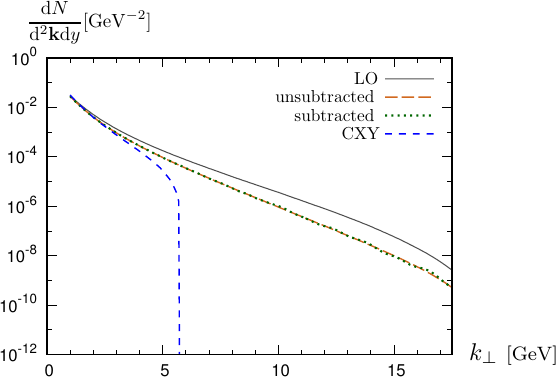}\\(a)\vspace{0cm}
\end{center}
\end{minipage}
\hspace{0.05\textwidth}
\begin{minipage}[b]{0.45\textwidth}
\begin{center}
\includegraphics[width=0.95\textwidth,angle=0]{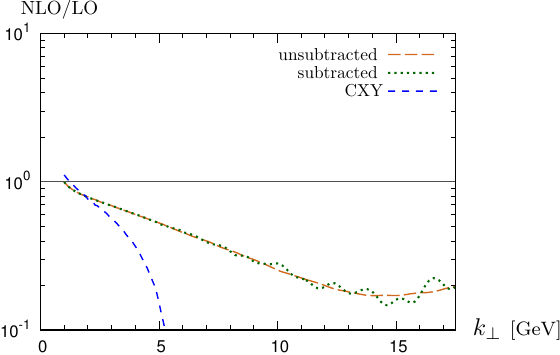}\\(b)\vspace{0cm}
\end{center}
\end{minipage}
\end{center}
\vspace{-0.25cm}
\caption{LO, NLO and $k_{\perp}$-factorization (CXY) results for the single inclusive ``quark production''. Adapted from \cite{Ducloue:2017mpb}.}
\label{fig-num}
\end{figure}

We repeat that, while the ``unsubtracted'' and ``subtracted'' expressions \eqref{dnnlo} and \eqref{dnnlo2} are mathematically identical, \eqn{dnnlo3} is not equivalent with them. Due to the approximations performed, it mistreats the kinematics at small-$x$ and leads to a negative cross section. Indeed, all this is confirmed in the numerical evaluation of all the aforementioned formulas for the cross section as shown in Fig.~\ref{fig-num}. We also note that, even though \eqn{dnnlo} and \eqn{dnnlo2} agree very well with each other, the subtracted expression exhibits some fluctuations for high $k_{\perp}$ as a result of the large reshuffling. One sees that the NLO correction is negative and with a large but under control magnitude: around $50\%$ at $k_{\perp}\sim 5$ GeV \cite{Ducloue:2017mpb}.

The discussion so far has been restricted to using a QCD coupling fixed to a moderately small value. It is not trivial at all to extend \eqn{dnnlo} with running coupling included. The dependence on $\abar$ there is in two places: the implicit BK-dependence in $\mcal{S}_{q\bar{q}g}\big(\bk,X(x)\big)$ and the explicit impact factor one in front of the integration. Regarding the former, that is the coupling along the small-$x$ ladder, one first solves the BK-equation in coordinate space with the smallest dipole prescription \cite{Iancu:2015joa}, i.e.~using $\abar(r_{\rm min})$ with $r_{\rm min}$ the size of the smallest dipole among the three involved in the splitting process in \eqn{bk} and then one transforms the results to momentum space. Regarding the latter, that is the coupling associated with the primary gluon, it is natural to use $\abar(k_{\perp})$ \cite{Iancu:2016vyg}, since $k_{\perp}$ is the momentum of the ``measured'' outgoing quark, and all this is leading again to well-defined results. In order to have a more homogeneous description, one may try to move the coupling inside the integration over the transverse coordinates \cite{Ducloue:2017mpb}. Without going into much detail here, we find that when the primary gluon has a moderate $x$ and is soft in the transverse space, the scale in the coupling must be set by the ``daughter dipole'' \cite{Ducloue:2017dit}, which is the largest one in the kinematics under consideration. A smallest dipole choice here would lead to heavily unstable results \cite{Ducloue:2017mpb}, since it generates a fake potential leading to unphysical contributions to the scattering \cite{Ducloue:2017dit}. Moreover, for the NLO terms in the $C_{\rm F}$-sector a coordinate space prescription for the coupling either cannot be implemented, or leaves uncanceled some spurious longitudinal logarithms \cite{Ducloue:2017dit}, so eventually $\abar(k_{\perp})$ is the only viable choice.  

\section{Conclusion}
\label{sec-conc}

Single inclusive particle production at forward rapidities in proton-nucleus collisions is a process suited to study gluon saturation. Early results at NLO in the CGC effective theory led to a negative cross section and here we have shown that this was the result of approximations which were necessary to arrive at an expression consistent with $k_{\perp}$-factorization. Thus, we have developed a generalized factorization procedure (which involves a longitudinally dependent kernel and is non local in rapidity) that gives by definition a well-defined cross section \cite{Iancu:2016vyg}. As expected, the NLO correction is negative and reduces the LO result by $\sim 50\%$ when the outgoing quark (that fragments to a hadron) has a transverse momentum of a few GeV.


\end{document}